\documentclass{iaus}
\usepackage{graphicx}

\title[IAU 262.~~Helium Enrichment in Globular Clusters] 
{Study of the Helium Enrichment\\ in Globular Clusters}

\author[Aldo A. R. Valcarce \& M\'arcio Catelan]   
{Aldo A. R. Valcarce
 \and M\'arcio Catelan}

\affiliation{Pontificia Universidad Cat\'olica de Chile, \\ 
Departamento de Astronom\'{i}a y Astrof\'{i}sica, \\ 
Av. Vicu\~na Mackenna 4860,
782-0436 Macul, Santiago, Chile \\ email: {\tt avalcarc@astro.puc.cl; mcatelan@astro.puc.cl} \\}

\pubyear{2009}
\volume{262}  
\pagerange{119--126}
\jname{Stellar Populations -- Planning for the Next Decade}
\editors{G. Bruzual \& S. Charlot, eds.}
\begin{document}

\maketitle

\begin{abstract}
Globular clusters (GCs) are spheroidal concentrations typically containing of the order of $10^5$ to $10^6$, predominantly old, stars. Historically, they have been considered as the closest counterparts of the idealized concept of ``simple stellar populations.'' However, some recent observations suggest than, at least in some GCs, some stars are present that have been formed with material processed by a previous generation of stars. In this sense, it has also been suggested that such material might be enriched in helium, and that blue horizontal branch stars in some GCs should accordingly be the natural progeny of such helium-enhanced stars. In this contribution we show that, at least in the case of M3 (NGC~5272), the suggested level of helium enrichment is not supported by the available, high-precision observations.

\end{abstract}

\firstsection 
\section{Introduction}
Globular clusters (GCs) have long been thought to provide one of the closest approximations to the idealized concept of ``simple stellar populations,'' with all stars in a given GC having closely the same distance, age, and chemical composition. However, our understanding of these objects is quickly changing, thanks not only to measurements of chemical inhomogeneities, but also to improved color-magnitud diagrams (CMDs), which show that at least some GCs present multiple populations, presumably indicative of multiple formation episodes. 

Besides the classical cases of $\omega$~Cen (e.g., \cite{Vetal07}) and M54 (e.g., Siegel et al. 2007), which host multiple red giant branches (RGBs), subgiant branches (SGBs), and main sequences (MSs), a triple MS has recently been identified in NGC~2808 (Piotto et al. 2007), whereas a double SGB has also been found in 47~Tuc (\cite{Aetal09}), M22 (\cite{Metal09}), NGC 1851 (\cite{Metal08}), NGC 6388 (\cite{P08}), and NGC 5286 (Piotto et al. 2009). Several of these CMD peculiarities often go hand-in-hand with multimodalities along the horizontal branch (HB). 

As a rule, different formation episodes inside the clusters have been suggested as the explanation for this phenomenon. However, depending on the specific case considered, different chemical enrichment scenarios have been advocated, including a spread in the CNO elements (\cite{Cass08}, \cite{Pietri09}) or/and various populations with different initial He abundances. As to the latter, a helium abundance of 40\% by mass has been found to be required in order to reproduce the bluest MSs in $\omega$~Cen and NGC~2808 (\cite{Petal05}, \cite[2007]{Petal07}).

In addition to this, it has also been recently suggested that the spread in colors seen along the HB in GCs should be due to a spread in $Y$ within these objects, in which case helium abundance enhancements would be the rule, rather than the exception, among GCs (e.g., \cite{CD07}, \cite{DC08}). In this sense, these authors have found that the helium abundance among M3's blue HB stars should be enhanced, with respect to M3's red HB population, by at least $\Delta Y = 0.02$. The main purpose of this contribution is to test this key prediction (see also \cite{Cetal09} for more details).

\begin{figure}[hu]
  \begin{center}
    \begin{minipage}[t]{0.5\linewidth}
      \raisebox{-1.5cm}{\includegraphics[height=2.5in]{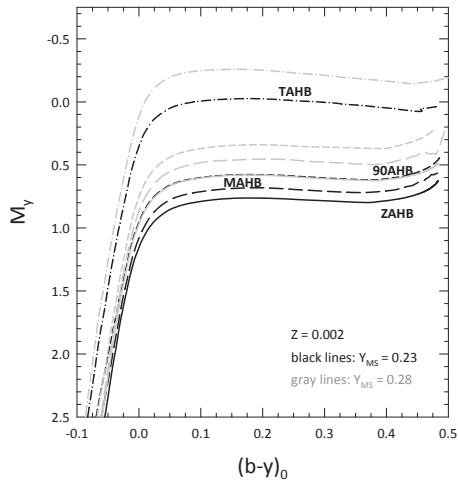}}
    \end{minipage}\hfill
    \begin{minipage}[t]{0.475\linewidth}
      \caption{Comparison between the equivalent evolutionary sequences (EESs) for a fixed metallicity and two different helium abundances (black lines: $Y = 0.23$; gray lines: $Y = 0.28$).}\label{fig1}
    \end{minipage}
  \end{center}
\end{figure}

\firstsection
\section{Is there a Helium-Enhanced Population in M3?}

To answer this question, we compared observational data for stars in the HB phase with zero age HB (ZAHB) models for different helium abundances.

\subsection{Observational Data and Theoretical Models}
Very briefly, we used the high-precision Str\"omgren photometry for the cluster, obtained with the Nordic Optical Telescope (\cite{Getal98}, \cite{Getal99}).

Theoretical models are from \cite{Cetal98} and \cite{SC98}, with heavy-element abundances $Z =0.0005$, 0.001, and 0.002, and He abundances $Y = 0.23$, $0.28$, and $0.33$. This nicely brackets the possible metallicity range for M3, which in the \cite{ZW84} scale and in the \cite{CG97} scale have ${\rm [Fe/H]} =-1.57$ and ${\rm [Fe/H]} = -1.34$, respectively. Taking into account an $[\alpha/{\rm Fe}]=+0.27$ for the cluster (\cite{C96}), these values imply a $Z =8.3\times 10^{-3}$ and $Z =1.4\times 10^{-3}$, respectively, based on the \cite{Setal93} $Z-{\rm [Fe/H]}-[\alpha/{\rm Fe}]$ relation. Model sequences from \cite{vdbea06} for a $Y = 0.236$ and $Z = 8.5 \times 10^{-4}$ have also been used, for comparison. 

To transfer the evolutionary tracks to the observational plane, we used the color transformations and bolometric correction tables from \cite{Clemetal04}.

\begin{figure}[hu]
\begin{center}
 \includegraphics[width=5.2in,height=6.025in]{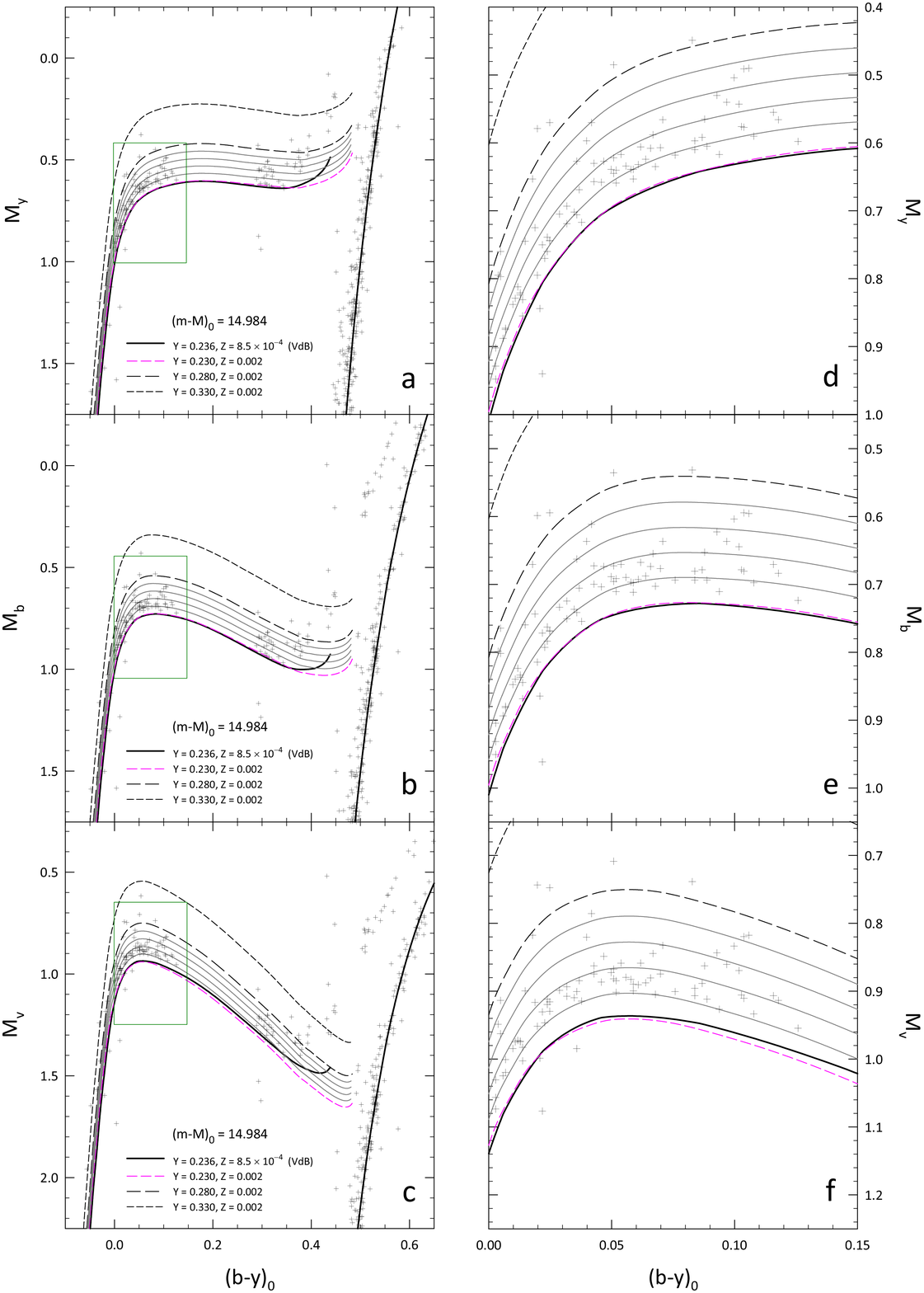} 
 \caption{Comparison between the empirical data for M3 and theoretical ZAHBs for the several different $Y$ values indicated. The thin solid lines indicate interpolated ZAHB loci for $Y$ values between 0.24 and 0.27, in intervals of 0.01. Panels d through f present expanded views of the rectangular box around the blue HB shown in panels a through c, respectively.}
   \label{fig2}
\end{center}
\end{figure}

\subsection{Signatures of Varying He Abundances}

To determine how different the HB morphology may become when the He abundance is changed, we used the following equivalent evolutionary sequences (EESs) for this phase: i)~The ZAHB; ii)~The terminal-age HB (TAHB), or He exhaustion locus; iii)~The middle-age HB (MAHB); iv)~The 90\%-age HB (90AHB). The latter two are the position corresponding to HB stars that have completed $50\%$ and $90\%$ of their total evolutionary history along the HB, respectively. Furthermore, the MAHB could be also assumed as the average position occupied by all HB stars, if there is a continuous star supply from the RGB. 

Figure~\ref{fig1} shows how these EESs change when $Y$ changes. Clearly, the luminosity of each EES increases very substantially for larger $Y$ values. Therefore, differences in the He abundance between red and blue HB stars in a given cluster (e.g., \cite{CD07}, \cite{DC08}) should imply a difference in magnitude between red and blue HB stars.

\subsection{Results}

We compare the empirical M3 CMD with ZAHBs for different He abundaces in Figure~\ref{fig2}. To produce these plots, we corrected the CMD data for reddening using a $E(B-V)=0.01$ (\cite{Harris96}); extinction coefficients for the Str\"omgren system were taken from \cite{CC08}. In these plots, the red HB of the cluster was fit to the ZAHB locus for a canonical $Y = 0.23$. 

Clearly, when the red HB of the cluster is forced to match ZAHB models with the canonical He abundance, the blue HB stars of the cluster are found to also agree with the same ZAHB locus, without the need for He enhancement. More specifically, our comparison shows that an increase in $Y$ by more than $0.01$ among the blue HB stars (with respect to the red HB stars in the same cluster) would not be compatible with the data. This strongly suggests that the level of He enhancement is most likely less than 0.01 among the cool blue HB stars in M3.

In the future, we will apply similar tests to other GCs which have been suggested to harbor He-enhanced populations, thus providing important constraints on the He-enhancement scenario. 

\firstsection
\section*{Acknowledgement}
We thank F. Grundahl, A.V. Sweigart, and C. Cort\'es for helpful discussions. Support for A.V. is provided by IAU, CONICYT, SOCHIAS, MECESUP, and ALMA. Support for M.C. is provided by Proyecto Basal PFB-06/2007, by FONDAP Centro de Astrof\'{i}sica 15010003, by Proyecto FONDECYT Regular \#1071002, and by a John Simon Guggenheim Memorial Foundation Fellowship.

\end{document}